\documentclass[letter]{jpsj2} 

\title{Novel phase transition and the pressure effect in \\
YbFe$_2$Al$_{10}$-type Ce$T_2$Al$_{10}$ ($T$=Fe, Ru, Os)}

\author{Takashi \textsc{Nishioka}$^{1}$\thanks{E-mail: nisioka@kochi-u.ac.jp}, 
Yukihiro \textsc{Kawamura}$^{1}$, 
Tomoaki \textsc{Takesaka}$^{1}$, 
Riki \textsc{Kobayashi}$^{1}$, 
Harukazu \textsc{Kato}$^{1}$, 
Masahiro \textsc{Matsumura}$^{1}$, 
Kazuto \textsc{Kodama}$^{2}$, 
Kazuyuki \textsc{Matsubayashi}$^{3}$ 
and Yoshiya \textsc{Uwatoko}$^{3}$}

\inst{$^{1}$Graduate School of Integrated Arts and Sciences, Kochi University, Kochi 780-8520, Japan \\
$^2$Center for Advanced Marine Research, Kochi University, Nankoku 783-8502, Japan\\
$^3$Institute for Solid State Physics, University of Tokyo, Chiba 277-8581, Japan}

\abst{
We have succeeded in growing single crystals of orthorhombic Ce$T_2$Al$_{10}$ ($T$=Fe, Ru, Os) by Al self-flux method for the first time, and measured the electrical resistivity $\rho$ at pressures up to 8 GPa, the magnetic susceptibility $\chi$ and specific heat $C$ at ambient pressure. These results indicate that Ce$T_2$Al$_{10}$ belongs to the heavy fermion compounds. 
CeRu$_2$Al$_{10}$ and CeOs$_2$Al$_{10}$ show a similar phase transition at $T_0$ = 27.3 and 28.7 K, respectively. 
The temperature dependences of $\rho$, $\chi$ and $C$ in the ordered phases are well described by the thermally activated form, suggesting that partial gap opens over the Fermi surfaces below $T_0$. 
When pressure is applied to CeRu$_2$Al$_{10}$, $T_0$ disappears suddenly between 3 and 4 GPa, and CeRu$_2$Al$_{10}$ turns into a Kondo insulator, followed by a metal. 
The similarity of Ce$T_2$Al$_{10}$ under respective pressures suggests a scaling relation by some parameter controlling the unusual physics in these compounds. 
}

\kword{YbFe$_2$Al$_{10}$-type, CeRu$_2$Al$_{10}$, CeFe$_2$Al$_{10}$, CeOs$_2$Al$_{10}$, electrical resistivity, magnetic susceptibility, specific heat, pressure, Kondo semiconductor, heavy fermion, CDW, cage structure}

\begin{document}
\maketitle

Magnetic properties of Ce compounds are relatively well described by the Doniach's phase diagram, i.e., the competition between Kondo and RKKY interactions. 
Although both interactions are attributed to the exchange coupling $J_\mathrm{cf}$ between the 4$f$ and conduction electrons, Kondo effect is one-ion effect with non-magnetic ground state, whereas RKKY interaction is two-ion effect with magnetic one. 
Even if RKKY interaction is generally a long range interaction, it decreases as $1/d^3_{\mbox{\scriptsize Ce-Ce}}$, where $d_{\mbox{\scriptsize Ce-Ce}}$ is the nearest distance between two Ce ions, and therefore the magnetic ordering temperature ($T_\mathrm{M}$) of the compound with large $d_{\mbox{\scriptsize Ce-Ce}}$ is expected to be suppressed. 
In fact, $T_\mathrm{M}$ of most of the Ce compounds in which $d_{\mbox{\scriptsize Ce-Ce}}$ exceeds 5 \AA \ are below 2 K.\cite{ser} When $J_\mathrm{cf}$ is strong enough, valence fluctuation behavior appears such as CeBe$_{13}$.\cite{wil} Therefore, Ce compounds with large $d_{\mbox{\scriptsize Ce-Ce}}$ have been simply considered to exhibit either localized magnetism with low $T_\mathrm{M}$ or valence fluctuation.

Recently, Strydom has reported that orthorhombic YbFe$_2$Al$_{10}$-type CeRu$_2$Al$_{10}$ shows a phase transition at 27 K. He proposed that the transition is attributed to a magnetic ordering and the paramagnetic phase is a semiconductor.\cite{str} CeRu$_2$Al$_{10}$ is a cage structure compound, where Ce ion is surrounded by 16 Al and 4 Ru, and $d_{\mbox{\scriptsize Ce-Ce}}$ is about 5.2 \AA.\cite{thi,tur} 
His arguments, however, are based on only macroscopic properties of polycrystalline samples including impurity phases, it should not be concluded that the transition is magnetic. 
Furthermore, the ordering temperature is surprisingly high to regard as a magnetic one. Indeed, GdRu$_2$Al$_{10}$ orders magnetically at 16.5 K from our experiment, and if we simply apply the de Gennes law, the expected $T_\mathrm{M}$ of CeRu$_2$Al$_{10}$ is less than 0.1 K. 
In addition, the cell volume is smaller than the value expected from the lanthanoid contraction, suggesting CeRu$_2$Al$_{10}$ is located in the valence fluctuation regime. 
Moreover, Muro {\it et al}. recently have reported CeFe$_2$Al$_{10}$ with the same structure is a Kondo semiconductor.\cite{mur} 
Since these studies were performed on polycrystalline samples, research using single crystals is required to reveal the origin of the transition of CeRu$_2$Al$_{10}$ and the relationship between CeRu$_2$Al$_{10}$ and CeFe$_2$Al$_{10}$. 

The purpose of this study is to clarify the properties of the phase transition in CeRu$_2$Al$_{10}$ and the relation with isostructural Ce$T_2$Al$_{10}$ ($T$=Fe, Os). 
In order to achieve this, we have grown single crystals of Ce$T_2$Al$_{10}$ ($T$=Fe, Ru, Os), and performed the electrical resistivity $\rho$ under pressure, magnetic susceptibility $\chi$ and specific heat $C$ measurements. 
In this paper, we describe mainly CeRu$_2$Al$_{10}$ and use $T_0$ as the transition temperature in order to distinguish from $T_\mathrm{M}$.

Single crystals were grown by Al self-flux method. 
The detail is described in our previous paper.\cite{tuk2} X-ray powder diffraction confirmed single phase for all samples and the crystallographic axes were determined using back reflection Laue method. 
The lattice constants were $a$=9.158 \AA, $b$=10.266 \AA, $c$=9.119 \AA \ for CeRu$_2$Al$_{10}$, $a$=9.009 \AA, $b$=10.227 \AA, $c$=9.076 \AA \ for CeFe$_2$Al$_{10}$, and $a$=9.164 \AA, $b$=10.253 \AA, $c$=9.137 \AA \ for CeOs$_2$Al$_{10}$. 
Their cell volumes deviate from the lanthanoid contraction. 
The electrical resistivity was measured using a standard ac or dc four-probe method. 
The magnetization was measured using MPMS. The specific heat was measured using PPMS. 
The low pressure ($\sim$1.5 GPa) and high pressure (2$\sim$8 GPa) were generated by a CuBe piston cylinder device and a cubic anvil device, respectively.

%
%
\begin{figure}[h]
\begin{center}
\includegraphics[width=0.94\textwidth]{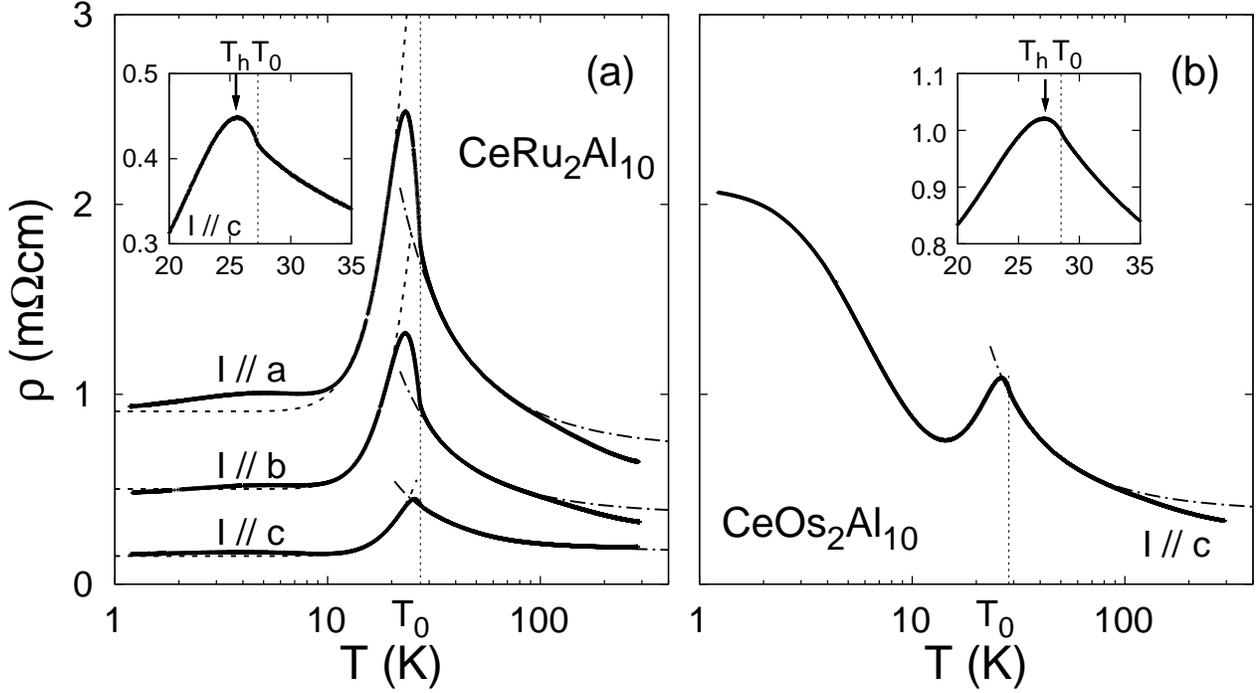}
\end{center}
\caption{
$\rho$ vs $T$ plot in a semi-logarithmic scale of CeRu$_2$Al$_{10}$ along the three principle axes (a) and CeOs$_2$Al$_{10}$ along the $c$-axis (b). 
The dash-dotted curves are fits to the thermally activated form $A_0\exp(\Delta_\mathrm{T}/2T)$, where $\Delta_\mathrm{T}$=47.3 K ($a$-axis), 48.9 K ($b$-axis), 47.8 K ($c$-axis) for CeRu$_2$Al$_{10}$; $\Delta_\mathrm{T}$=55.5 K ($c$-axis) for CeOs$_2$Al$_{10}$. 
The dotted curves in (a) are fits by an antiferromagnetic magnon gap form (see text). The insets show an enlarged scale at around $T_0$. The down-arrows indicate the peak temperatures.
}
\label{f1}
\end{figure}
Figure 1(a) shows $\rho$($T$) of CeRu$_2$Al$_{10}$ single crystal. The overall shape of $\rho$($T$) along each principle axis resembles apart from the difference of their absolute values. 
As temperature decreases from room temperature, $\rho$ increases monotonically and shows an abrupt increase at $T_0$=27.3 K, passes through a hump and then decreases. 
We can see a slight decrease for all axes below $\sim$4 K. 
Although the hump along the $c$-axis is less visible in the main panel, it certainly exists as shown in the inset. The hump temperatures $T_\mathrm{h}$ along the $a$-, $b$- and $c$-axes are 23.2, 23.2 and 25.6 K, respectively. 

As shown by the dash-dotted curves, the temperature dependences above $T_0$ cannot be fitted completely by the thermally activated form $A_0\exp(\Delta_\mathrm{T}/2T)$ pointed out by Strydom.\cite{str}. 
The obtained $\Delta_\mathrm{T}$ for narrow fitting range are about 40 K, twice as large as the value obtained by Strydom.\cite{str} 
1/$T_1$ measurements of our $^{27}$Al NQR study does not indicate any gap-type temperature dependence above $T_0$.\cite{mat} 
Lying on straight lines in a  semi-logarithmic plot of $\rho(T)$ above $\sim$100 K might be due to Kondo effect. As indicated by the dotted curves, $\rho(T)$ for all axes below $T_\mathrm{h}$ are relatively well described by the expression as used by Strydom:\cite{str}
$
 \rho(T)=\rho(0)+aT(1+2T/\Delta_\rho)\exp(-\Delta_\rho/T)+AT^2
$, 
which is expected for an energy gap antiferromagnetic (AF) magnon with a Fermi-liquid $T^2$ and a residual resistivity term. The least square fit gives $\Delta_\rho$=32.9, 40.7 and 41.3 K for $a$-, $b$- and $c$-axis, respectively, and the $A$ term is negligible. 
It is peculiar that $\rho(T)$ for all axes can be fitted by this equation, though $\rho(T)$ should follow it for only one direction in the case of AF magnon gap. 
Since the evidence of AF ordering is not detected in NQR measurement,\cite{mat} gapped magnon excitation is not applied in this system.
The origin of the gap is probably not due to AF magnon. 
The abrupt increase in $\rho$ might be due to substantial reduction of conduction electrons by partial gap opening over the Fermi surface below $T_0$, and the metallic behavior at low temperatures comes from the remaining conduction electrons. 
Figure 1(b) shows $\rho(T)$ of CeOs$_2$Al$_{10}$ along the $a$-axis. 
The shape of $\rho(T)$ is very similar to $\rho_\mathrm{a}(T)$ except below $\sim$15 K. 
CeOs$_2$Al$_{10}$ also shows a small increase below $T_0$=28.7 K and a small hump at $T_\mathrm{h}$=27.0 K. 
The increase in $\rho$ below $\sim$15 K may be attributed to larger gapped region over the Fermi surface compared to that of CeRu$_2$Al$_{10}$. 

%
%
\begin{figure}[h]
\begin{center}
\includegraphics[width=0.94\textwidth]{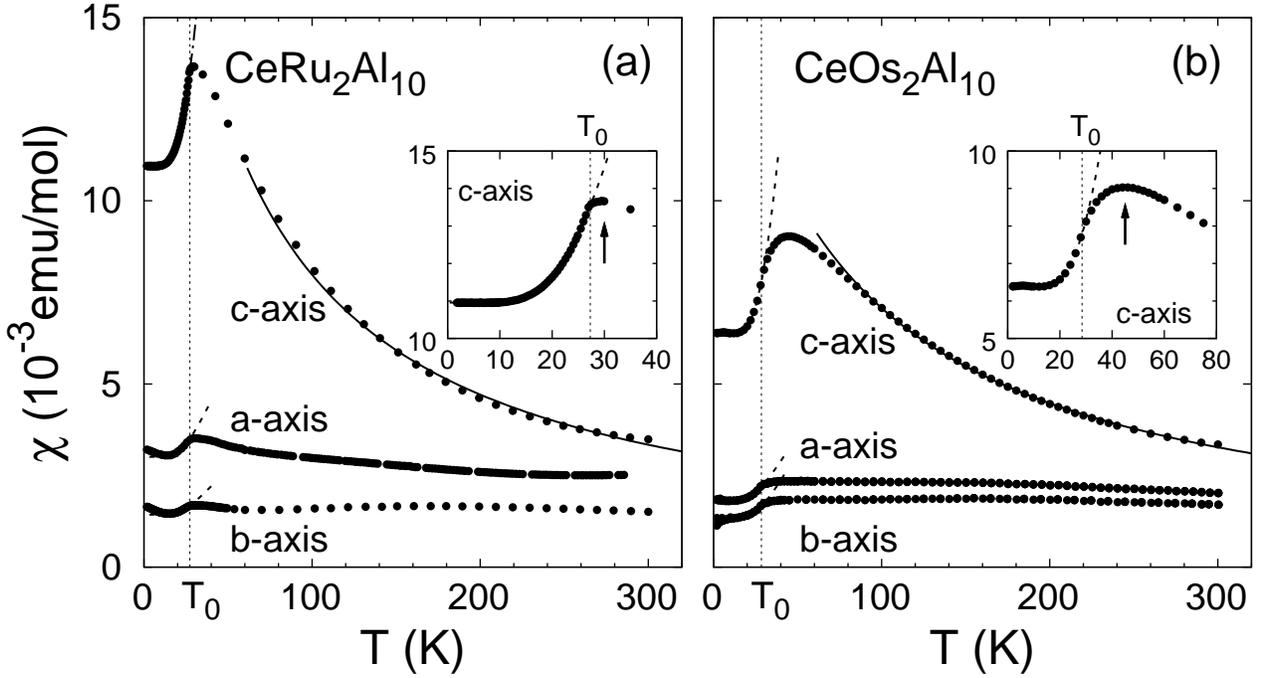}
\end{center}
\caption{$\chi$ vs $T$ plot for CeRu$_2$Al$_{10}$ (a) and CeOs$_2$Al$_{10}$ (b). The applied fields are 5 T and 1 T for CeRu$_2$Al$_{10}$ and CeOs$_2$Al$_{10}$, respectively. 
The solid lines indicate a least square fit to the Curie-Weiss law. The inset shows an enlarged scale at around $T_0$ for the $c$-axis. The up-arrows indicate peak temperatures. The dotted curves are a least square fit to the 
$\chi$=$\chi_0$+$A_1\exp(-\Delta_\chi/T)$, where $\Delta_\chi$=96 (100) K for $a$-axis, 96 (114) K for $b$-axis, 101 (137) K for $c$-axis for CeRu$_2$Al$_{10}$ (CeOs$_2$Al$_{10}$).}
\label{f2}
\end{figure}
Figures 2(a) and 2(b) show $\chi(T)$ of CeRu$_2$Al$_{10}$ and CeOs$_2$Al$_{10}$, respectively. 
The $\chi(T)$ indicates uniaxial magnetic anisotropy with an easy $c$-axis. 
The $\chi$ along the $c$-axis for each compound nearly follows the Curie-Weiss law with the effective magnetic moment $\mu_\mathrm{eff}$=3.0 (3.1) $\mu_\mathrm{B}$/Ce, the paramagnetic Curie temperature $\Theta_\mathrm{p}$=$-44$ ($-72$) K for CeRu$_2$Al$_{10}$ (CeOs$_2$Al$_{10}$).
Somewhat large $\mu_\mathrm{eff}$ compared to the theoretical Ce$^{3+}$ ion value (2.54 $\mu_\mathrm{B}$/Ce) and large negative $\Theta_\mathrm{p}$ are characteristics of valence fluctuation materials. 
The $\chi$ along the other axes are almost temperature independent. 
The large magnetic anisotropy should be attributed to the crystalline electric fields. 
As shown by the up-arrow in the inset, $\chi$ of CeRu$_2$Al$_{10}$ (CeOs$_2$Al$_{10}$) exhibits a broad peak at 30 (45) K, and decreases suddenly below $T_0$. 
The coincidence of the sudden decrease in $\chi$ and the abrupt increase in $\rho$ indicates some phase transition taking place at $T_0$. 
The previous study using a polycrystalline sample exhibits an upturn in $\chi$ at low temperatures due to impurities,\cite{str} while our single crystals do not show such behavior. 
The sudden decrease below $T_0$ is also observed for $b$- and $c$-axis. 
As shown by the dotted curves, $\chi(T)$ follows an exponential function $\chi$=$\chi_0$+$A_1\exp(-\Delta_\chi/T)$ below $T_0$, where $\Delta_\chi$=96$\sim$101 K (100$\sim$137 K) for CeRu$_2$Al$_{10}$ (CeOs$_2$Al$_{10}$). 
$\Delta_\chi$ for CeRu$_2$Al$_{10}$ is about twice as large as $\Delta_\rho$. 
Similar temperature dependence is also observed for the other axes. 

%
%
\begin{figure}[h]
\begin{center}
\includegraphics[width=0.94\textwidth]{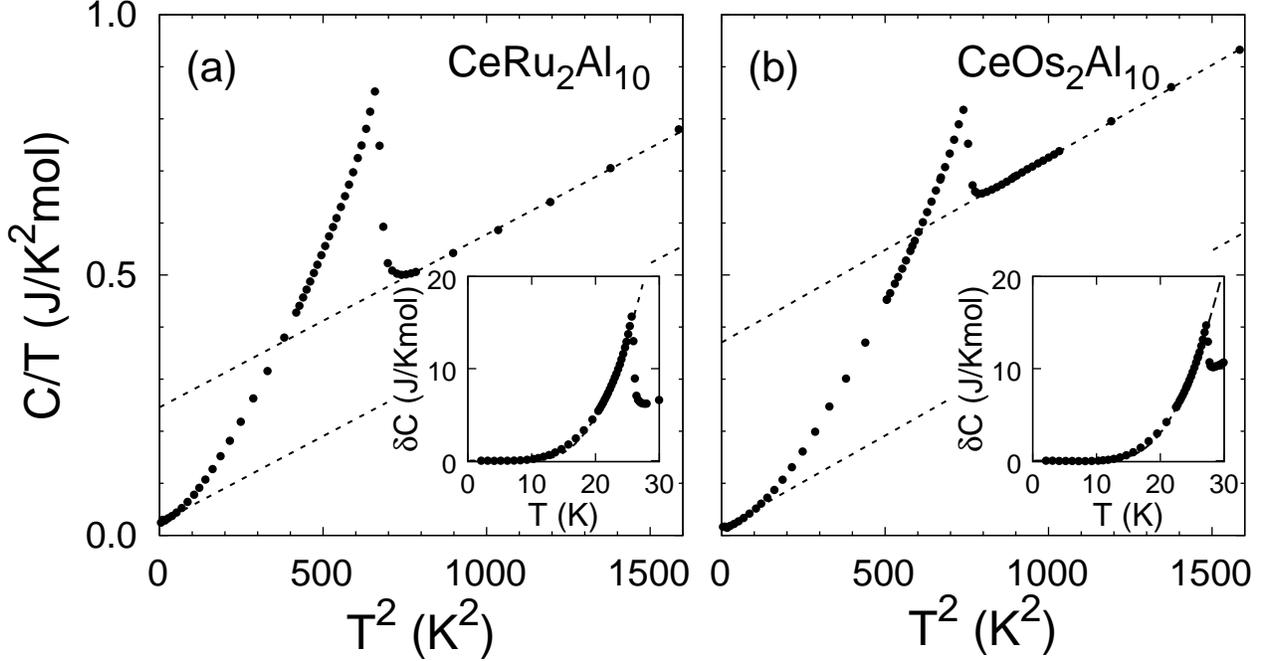}
\end{center}
\caption{$C/T$ vs $T^2$ plots for CeRu$_2$Al$_{10}$ (a) and CeOs$_2$Al$_{10}$ (b). The insets are plotted as $\delta C$ (=$C$$-\gamma_0T$$-\beta T^3$) vs $T$. The upper dashed upper straight lines are least square fits of $C/T$=$\gamma_1$+$\beta T^2$ in the temperature range 800$<$$T^2$$<$2000, where $\gamma_1$=246 (370) mJ/K$^2$mol, $\beta$=0.332 (0.355) mJ/K$^4$mol for CeRu$_2$Al$_{10}$ (CeOs$_2$Al$_{10}$). The calculated Debye temperatures $\Theta_\mathrm{D}$ is 423 (414) K for CeRu$_2$Al$_{10}$ (CeOs$_2$Al$_{10}$). The lower dashed straight lines are parallel displacement the above fitting lines. The dashed curves in the insets a fit of $\delta C$=$A_2\exp(-\Delta_C/T)$, where $\gamma_0$=24.5 (14.0) mJ/K$^2$mol, $\Delta_C$=102 (117) K for CeRu$_2$Al$_{10}$ (CeOs$_2$Al$_{10}$).}
\label{f3}
\end{figure}
Figures 3(a) and 3(b) show plots of $C/T$ vs $T^2$ for CeRu$_2$Al$_{10}$ and CeOs$_2$Al$_{10}$, respectively. The data points above $T_0$ lie on a straight line for each compound, indicating $C$=$\gamma_1T$+$\beta T^3$ with $\gamma_1$=246 (370) mJ/K$^2$mol for CeRu$_2$Al$_{10}$ (CeOs$_2$Al$_{10}$).  
The large linear specific coefficients $\gamma_1$ indicate both compounds belong to the heavy fermion (HF) compounds, while the values are reduced to $\gamma_0$=24.5 (14.0) mJ/K$^2$mol for CeRu$_2$Al$_{10}$ (CeOs$_2$Al$_{10}$). 
We can see a sharp peak in $C$ at 25.7 (27.2) K for CeRu$_2$Al$_{10}$ (CeOs$_2$Al$_{10}$). 
The shape below the peak temperature of $C$ is reminiscent of a second-order BCS-type mean-field transition. 
The insets are plots of $\delta C$ (=$C$$-\gamma_0 T$$-\beta T^3$) vs $T$ with a least square fit to the formula $\delta C$=$A_2\exp(-\Delta_C/T)$ denoted by the dotted lines, which give $\Delta_C$=102 (117) K for CeRu$_2$Al$_{10}$ (CeOs$_2$Al$_{10}$). 
The values of $\Delta_C$, which are indicated in the caption in Fig. 3, are very close to $\Delta_\chi$. 
From the ratio of $\gamma_1$ to $\gamma_0$, the transition removes 91\% (97\%) of the Fermi surface for CeRu$_2$Al$_{10}$ (CeOs$_2$Al$_{10}$). The very small remaining Fermi surface in CeOs$_2$Al$_{10}$ is probably the cause of low temperature increase in $\rho$. 
The value of the jump $\Delta C$ at $T_0$ is 8.35 (3.84) J/Kmol for CeRu$_2$Al$_{10}$ (CeOs$_2$Al$_{10}$), which gives $\Delta C/\gamma T_0$=1.24 (0.38). 
The relation between energy gaps and $T_0$ are $\Delta_C$=3.74 $T_0$ (4.11 $T_0$) for CeRu$_2$Al$_{10}$ (CeOs$_2$Al$_{10}$) and $\Delta_\rho$=1.2$\sim$1.5 $T_0$ for CeRu$_2$Al$_{10}$. 
The estimated magnetic entropy of CeRu$_2$Al$_{10}$ by subtracting $C$ of LaRu$_2$Al$_{10}$ is $\sim$0.7$R\ln 2$ at $T_0$ and $\sim$$R\ln 2$ at 100 K, indicating the ground state is doublet, so that a multi-pole ordering is completely excluded.

The above experimental results suggest that both CeRu$_2$Al$_{10}$ and CeOs$_2$Al$_{10}$ show a similar phase transition which opens partial gap over the Fermi surface below $T_0$. 
The temperature dependence of 1/$T_1$ of $^{27}$Al NQR measurement on CeRu$_2$Al$_{10}$ also indicates gap-like behavior. 
Furthermore, the temperature dependence of $^{27}$Al NQR spectra clearly indicate the transition is not attributed to magnetic origin but structural one making inequivalent Al site double.\cite{mat} 
Therefore, one possible candidate for the transition at $T_0$ is CDW. 
In general, CDW transition is based on one-dimensional physics and if it occurs in three dimensional materials, the existence of nesting vector should be expected. 
Since Ce$T_2$Al$_{10}$ is three-dimensional material as well as orthorhombic, it is peculiar that gap opens more than 90\% by the transition. 
If CDW transition certainly occurs, the CDW instability predicted by Overhauser in 1962 may be realized.\cite{ove}.

%
%
\begin{figure}[h]
\begin{center}
\includegraphics[width=0.94\textwidth]{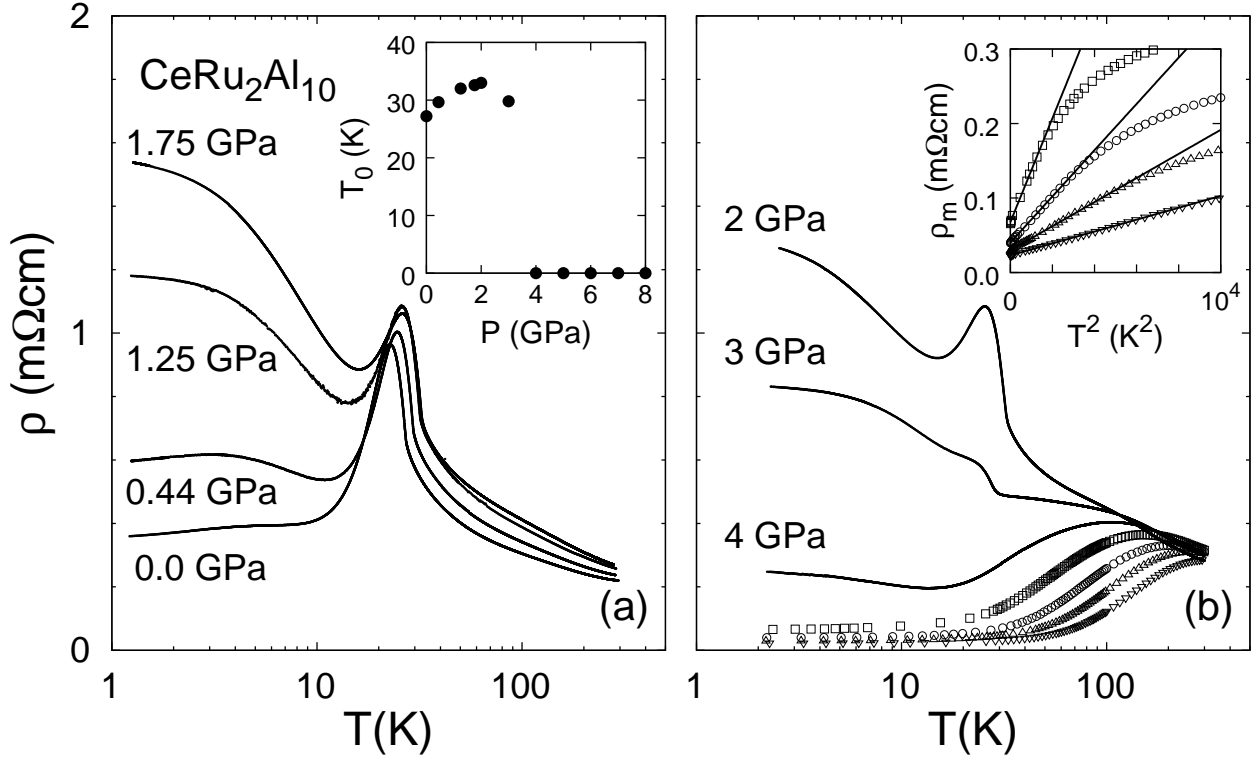}
\end{center}
\caption{$\rho$ vs $T$ plot in semi-logarithmic scale under pressure, where the pressure below 1.5 GPa is generated by a piston cylinder clamp (a) and the hydrostatic pressure above 2 GPa is generated by a cubic anvil device (b). The symbols used in (b) are $\Box$: 5 GPa, $\bigcirc$: 6 GPa, $\bigtriangleup$: 7 GPa, $\bigtriangledown$: 8 GPa. 
The inset in (a) is a plot of $T_0$ vs $P$. 
The inset in (b) is a plot of $\rho_\mathrm{m}$ vs $T^2$ at pressures between 5 and 8 GPa. The $\rho_\mathrm{m}$ was obtained by subtracting $\rho$ of LaRu$_2$Al$_{10}$.}
\label{f4}
\end{figure}
Figure 4 shows $\rho(T)$ along the $b$-axis of CeRu$_2$Al$_{10}$ under pressure. The most prominent variation in Fig. 4(a) is the behavior below $\sim$10 K. 
The metallic decrease at low temperatures initially turns into semiconducting-like increase by applying pressure. 
Further increase in pressure gradually weakens semiconducting increase and finally metallic decrease appears again above 5 GPa. 
$T_0$ increases gradually and reaches a maximum value at around 2 GPa, and then disappears suddenly between 3 and 4 GPa as pressure increases. 
The hump appeared below $T_0$ is little affected by pressure below 2 GPa, but shrinks at 3 GPa, and finally collapses with $T_0$. 
At 4 GPa, where $T_0$ disappears, a broad peak emerges at around 80 K. 
The peak temperature increases roughly in proportion to pressure, and reaches to room temperature at 8 GPa. 
The shape of $\rho(T)$ at 4 GPa, which is very similar to that of CeFe$_2$Al$_{10}$,\cite{mur} suggests a Kondo semiconductor. 
When pressure is further applied, semiconducting increase at low temperatures disappears and changes into metallic behavior above 5 GPa. 
The $\rho(T)$ above 5 GPa is very similar to the canonical HF system, i.e., $\rho$ increases in proportion to $-\log T$ at high temperatures and decreases in proportion to $T^2$ at low temperatures through a broad maximum. 

As shown in the inset in Fig. 4(b), $\rho_\mathrm{m}$ is in proportion to $T^2$ over wide temperature range. The $T^2$ coefficient at 5 GPa is $\sim$0.7 $\mu \Omega$cm/K$^2$. 
If Kadowaki-Woods relation holds,\cite{kad} the electronic specific coefficient would be expected to as $\sim$84 mJ/K$^2$mol for HF state. Usual HF state is realized in a region where a magnetic order is suppressed. 
Whereas, this HF state in this compound is realized due to the suppression of phase transition with non-magnetic origin. 
Superconductivity sometimes appears at around quantum critical point where a phase transition disappears.
We have tried to measure the $\rho(T)$ down to 0.7 K at pressures of 4.6 and 5.9 GPa using a Palm cubic-anvil device, but no superconducting transition was detected.

%
%
\begin{figure}[h]
\begin{center}
\includegraphics[width=0.94\textwidth]{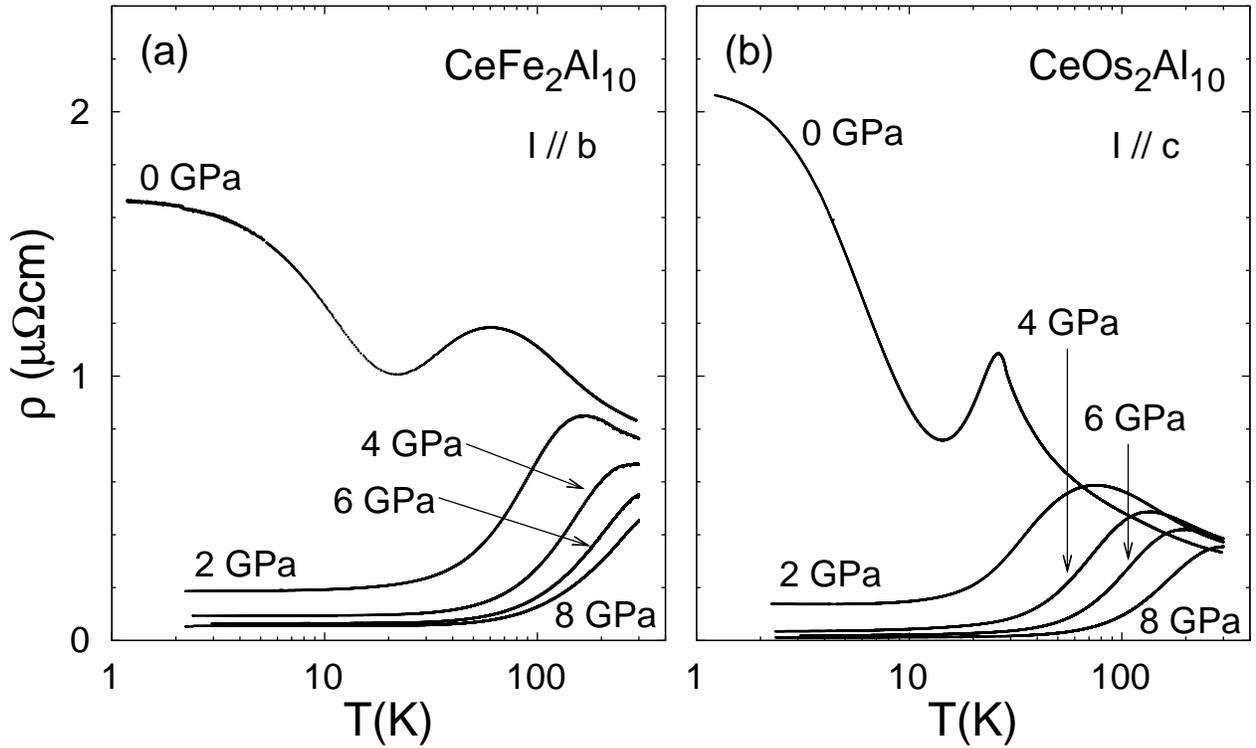}
\end{center}
\caption{$\rho$ vs $T$ plot in a semi-logarithmic scale at hydrostatic pressures up to 8 GPa for CeFe$_2$Al$_{10}$ along the $b$-axis (a) and CeOs$_2$Al$_{10}$ along the $c$-axis (b)}
\label{f5}
\end{figure}
Figures 5(a) and 5(b) show $\rho(T)$ at several pressures of CeFe$_2$Al$_{10}$ and CeOs$_2$Al$_{10}$, respectively. The $\rho(T)$ of CeFe$_2$Al$_{10}$ is very similar to a polycrystalline result.\cite{mur} 
As mentioned above, $\rho(T)$ in CeFe$_2$Al$_{10}$ is similar to that at 4 GPa in CeRu$_2$Al$_{10}$, so that CeFe$_2$Al$_{10}$ is expected to turn into a metal by applying pressure.
In fact CeFe$_2$Al$_{10}$ turns into a metal at 2 GPa with a broad maximum at $\sim$100 K. 
The maximum moves to higher temperatures with pressure like as CeRu$_2$Al$_{10}$. 
The $\rho(T)$ of CeOs$_2$Al$_{10}$ is very similar to that at 1.5 GPa in CeRu$_2$Al$_{10}$. 
The smaller hump compared to CeRu$_2$Al$_{10}$ is due to the anisotropy. 
$T_0$ disappears and a broad maximum appears at around 100 K at 2 GPa and the maximum temperature increases with pressure. 
This is also expected from the pressure dependence of CeRu$_2$Al$_{10}$.
The detailed pressure experiment will be published in a forthcoming paper.

In summary, we have firstly succeeded to synthesize single crystals of Ce$T_2$Al$_{10}$ ($T$=Fe, Ru, Os) by Al self-flux method, and measured $\rho(T)$ under pressure, $\chi(T)$ and $C(T)$. These results indicate that Ce$T_2$Al$_{10}$ show Kondo behavior at high temperature. 
CeRu$_2$Al$_{10}$ (CeOs$_2$Al$_{10}$) exhibits a phase transition at $T_0$=27.3 (28.7) K. 
This transition is characterized by an abrupt increase in $\rho$, a sudden decrease in $\chi$ and BCS-type anomaly in $C$. 
The temperature dependences of these macroscopic data suggest that CDW-like transition opens a gap over a portion of the Fermi surface. 
$^{27}$Al NQR measurements also suggest gap formation below $T_0$ and this transition not being magnetic origin.\cite{mat} 
According to the high pressure $\rho$ measurements of CeRu$_2$Al$_{10}$, $T_0$ increases slightly and disappears suddenly between 3 and 4 GPa, and CeRu$_2$Al$_{10}$ turns to a Kondo semiconductor. 
As pressure increases further, CeRu$_2$Al$_{10}$ turns into a metal above 5 GPa. The large $T^2$ coefficient suggests the metallic state is HF. 
The shape of $\rho(T)$ at ambient pressure of CeOs$_2$Al$_{10}$ (CeFe$_2$Al$_{10}$) is quite similar to that at 1.5 (4.5) GPa, and their pressure dependences are what expected from the pressure dependence of $\rho(T)$ in CeRu$_2$Al$_{10}$. 
Therefore, $\rho(T)$ in Ce$T_2$Al$_{10}$ can be roughly scaled by some physical parameter depending on pressure.

\section*{Acknowledgment}
We would like to thank Professors H. Yasuoka, T. Takabatake, M Sera, K. Iida and Doctor Y. Muro for valuable discussions. 
We also thank Professor H. Yoshizawa for the use of PPMS, Mr. A. Yamada for the pressure experiment using a palm cubic anvil device, and Mr. Y. Ogane for experimental assistance. 
This work was partially supported by a Grant-in-Aid for Scientific Research (C) of the Ministry of Education, Culture, Sports, Science and Technology, Japan and the Comprehensive Support Programs for Creation of Regional Innovation of Japan Science and Technology Agency (JST).

\end{document}